\documentclass{PoS}
\usepackage{macros}
\usepackage{color}
\let\normalcolor\relax
\definecolor{mygreen}{rgb}{0,0.3,0}

\title{Toward Group Theory Operators for tmLQCD Hadrons}

\ShortTitle{Toward Group Theory Operators for tmLQCD Hadrons}

\author{Derek Harnett\\
        Department of Physics, University College of the Fraser Valley, Abbotsford, BC, Canada, V2S~7M8\\
        E-mail: \email{derek.harnett@ucfv.ca}}
\author{Randy Lewis and \speaker{Robert G. Petry}\\
        Department of Physics, University of Regina, Regina, SK, Canada,
                    S4S~0A2\\
        E-mail: \email{randy.lewis@uregina.ca},
        \email{rob.petry@uregina.ca}}

\abstract{Extraction of the mass spectrum from twisted mass
lattice QCD is facilitated by respecting the octahedral group of
rotations and accommodating the broken parity and
flavor symmetries of the theory.  In this work,
group theory meson operators adapted to these
constraints are constructed for the special case of
quark and antiquark fields at a common lattice site,
connected by extended gauge field paths.
}

\FullConference{XXIVth International Symposium on Lattice Field Theory\\
                July 23-28, 2006\\
                Tucson, Arizona, USA}
\begin{document}
\section{Introduction}
\subsection{Motivation}
Twisted mass lattice QCD (tmLQCD) offers an efficient mechanism for
eliminating unphysical zero modes thereby admitting calculations at
lighter quark masses\cite{Frezzotti:2000nk}. At maximal twist one has
$O(a)$ improvement\cite{Frezzotti:2003ni}. Lighter quark mass
calculations require improved statistics for hadron mass resolution.
This fact and the need to disentangle physical states with the same
quantum numbers may in part be accomplished through the creation of
more operators that represent the channel in question.  The creation
of more elaborate operators also allows for the study of hybrid and
exotic mesons.

The violation of parity by the twisted mass action causes many
operators to be twist angle dependent resulting in correlator
contamination by opposite parity states when not at maximal twist.
This suggests seeking a class of twist independent operators from
which the physical parity of an operator is readily deduced.

Operators with displaced quarks have been constructed using group
theoretical techniques but their usage necessarily requires the
calculation of quark propagators from multiple sources.  (See, for
example, Refs. \cite{Lacock:1996vy,Basak:2005aq,Basak:2005ir}.)
Consideration of the operators available through gluonic extension
alone therefore is calculationally expedient.

\subsection{Lattice Symmetry Group}
While parity ($P$) and charge conjugation ($C$) may be conserved by
lattice actions, the continuous rotational symmetry of nature is
broken and one requires operators adapted to the symmetry group of the
lattice.  For mesons this is the octahedral group $\Ogroup$ with $24$
elements. The direct product of the parity, charge conjugation, and
octahedral groups is denoted $\Ogroup^{PC}$.

For an operator adapted to the representation $\irrep^{PC}$, where
\mbox{$\irrep\in\{A_1,A_2,E,T_1,T_2\}$} is an irreducible
representation (irrep) of $\Ogroup$ and $P,C\in\{+,-\}$, one
identifies the possible physical states $J^{PC}$ to which it
corresponds using Table~\ref{tab:irrepcorrespondence} which shows the
number of copies $n^J_\irrep$ of irrep $\irrep$ to which the continuum
$SU(2)$ irrep $J$ subduces\cite{Johnson:1982yq}.
\begin{table}[h]
\begin{center}
$
\begin{array}{|r|rrrrr||r|rrrrr|} \hline
J & A_1 & A_2 & E & T_1 & T_2 &
J & A_1 & A_2 & E & T_1 & T_2 \\ \hline
0 & 1 & 0 & 0 & 0 & 0 &
3 & 0 & 1 & 0 & 1 & 1 \\
1 & 0 & 0 & 0 & 1 & 0 & 
4 & 1 & 0 & 1 & 1 & 1 \\
2 & 0 & 0 & 1 & 0 & 1 & 
5 & 0 & 0 & 1 & 2 & 1 \\ \hline
\end{array}
$
\end{center}
\caption{Subduction of continuum irreps to those of the lattice symmetry group.}
\label{tab:irrepcorrespondence}
\end{table}
An operator transforming as the irrep $E^{+-}$ could have spin content
\mbox{($J^{PC}=2^{+-},4^{+-},5^{+-},\ldots$)}. In theory one needs
many operators of each $\irrep^{PC}$ to resolve the many physical
spins $J^{PC}$ in the tower of states to which $\irrep^{PC}$
corresponds.

\subsection{Operator Building Blocks}
\begin{figure}[h]
\center{\includegraphics[width=.2\textwidth]{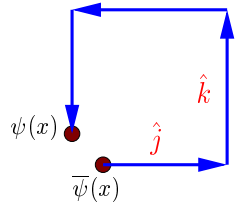}}
\caption{Diagrammatic representation of the building block operator.}
\label{fig:buildingblock}
\end{figure}

One may construct zero-momentum operators transforming as irreps of
$\Ogroup^{PC}$ from the space of operators spanned by
\begin{equation}
\label{eq:1}
M_{j,k,a,b}(t)=\sum_\mathbf{x}\spinorbar_a(x)
 U_j(x)U_k(x+\unit{j})U_{-j}(x+\unit{j}+\unit{k})U_{-k}(x+\unit{k}) \spinor_b(x)\ ,
\end{equation}
where $j,k = \pm 1,\pm 2,\pm 3; j \ne k$ and $a$ and $b$ are spinor
indices for a total of $24\times16=384$ operators.
See Figure~\ref{fig:buildingblock} for a diagrammatic representation. 
Superpositions of such operators for meson spectra analysis have been
suggested and used previously in special
cases\cite{Mandula:1983wc,Jorysz:1987qj,Lacock:1996vy,Bernard:1997ib}.
Evaluation of correlators constructed from operators of this form
requires only calculation of propagators from a single source.
Construction of operators transforming as irrep $\irrep^{PC}$ is
facilitated by observing that one can effectively consider the
transformation of the spinor and link paths independently and then
combine them via octahedral Clebsch-Gordan coefficients, as shown in
Sections \ref{sec:spin-contr-group}, \ref{sec:link-contr-group}, and
\ref{sec:total-repr-reduct}.
%

\section{Spinor Contribution to Group Structure}
\label{sec:spin-contr-group}
The contribution to the group structure due to spinor indices is
determined by the $16$ bilinears $\spinorbar\dirac
\spinor$, where $\dirac$ represents one of sixteen $4\times4$ matrices,
\mbox{$\{I,\gamma_5,\gamma_4,\gamma_4\gamma_5,
  \gamma_i,\gamma_i\gamma_5,\sigma_{4i},\epsilon_{ijk}\sigma_{jk}\}$}.
The first four bilinears are scalars while the last four three-index
objects are vectors under the rotation group in the Euclidean
continuum.

Parity ($P$) and charge conjugation ($C$) of the bilinears are
identified via
\begin{equation}
\begin{array}{ccc}
\mathcal{C}\spinor\mathcal{C}^\dagger = \left(\spinorbar C^\dagger\right)^T &
\hspace{1em} &
\mathcal{P}\spinor\mathcal{P}^\dagger = \gamma_4 \spinor \\
\mathcal{C}\spinorbar\mathcal{C}^\dagger = -\left(C \spinor\right)^T &
&
\mathcal{P}\spinorbar\mathcal{P}^\dagger = \spinorbar \gamma_4
\end{array}\ ,
\end{equation}
where $C$ is the matrix implementing charge conjugation.

To classify the bilinears into irreps of $\Ogroup$ one uses its
character table found in Table \ref{tab:ochartable}\cite{Johnson:1982yq}.
\begin{table}[h]
\begin{center}
$\begin{array}{|c|rrrrr|} \hline
\irrep\backslash\class & E & 3C_4^2 & 8C_3 & 6 C_4 & 6 C_2 \\ \hline
A_1 & 1 & 1 & 1 & 1 & 1 \\
A_2 & 1 & 1 & 1 & -1 & -1 \\
E & 2 & 2 & -1 & 0 & 0 \\
T_1 & 3 & -1 & 0 & 1 & -1 \\
T_2 & 3 & -1 & 0 & -1 & 1 \\ \hline
\end{array}
$
\end{center}
\caption{Character table giving $\character(\class)$ for octahedral group $\Ogroup$.}
\label{tab:ochartable}
\end{table}
Since the bilinears are scalars and vectors in the continuum it
follows that their respective spans are also invariant under
$\Ogroup$.  By inspection it is straightforward to show that the
character table for the scalar and vector bilinears is identical with
that of $A_1$ and $T_1$ respectively.  The multiplicity $n^J_\irrep$
of irrep $\irrep$ in the representation $J$ is given by
\begin{equation}
\label{eq:2}
n^J_\irrep=\frac{1}{\order{\Ogroup}} \sum_\class p_\class
\character^{(\irrep)}(\class)^*\character^{(J)}(\class)\ ,
\end{equation}
where $p_\class$ is the number of elements in class $\class$ and
$\order{\Ogroup}$ is the order of the octahedral
group\cite{Johnson:1982yq}.  It follows trivially by this formula that
the scalar and vector bilinears form the basis of $A_1$ and $T_1$
irreps respectively.  Furthermore it may be verified that the $i^{th}$
vector bilinear component transforms as the $i^{th}$ row for our
choice of matrix representation $\repmatrix{T_1}{R}$.

Hence the reduction of the spinor structure of our operators is given
in Table \ref{tab:spinorreduction}, where now each bilinear may be
classified uniquely by its irrep $\irrep^{PC}$, row~$\irrepi$, and
irrep multiplicity index~$\alpha$.
\begin{table}[h]
\begin{center}
$
\begin{array}{|c|c|c|c||c|c|c|c|} \hline
\dirac & \rule{0ex}{2.5ex}\irrep^{PC} & \irrepi & \irrepcopy &
\dirac & \rule{0ex}{2.5ex}\irrep^{PC} & \irrepi & \irrepcopy \\ \hline
\color{red}{I} & \rule{0ex}{2.5ex}\color{red}{A_1^{++}} & \color{red}{1} & \color{red}{1} &
\color{mygreen}{\gamma_i} & \rule{0ex}{2.5ex}\color{mygreen}{T_1^{--}} & \color{mygreen}{i} & \color{mygreen}{1} \\
\color{red}{\gamma_5} & \rule{0ex}{2.5ex}\color{red}{A_1^{-+}} & \color{red}{1} & \color{red}{1} &
\color{mygreen}{\gamma_i\gamma_5} & \rule{0ex}{2.5ex}\color{mygreen}{T_1^{++}} & \color{mygreen}{i} & \color{mygreen}{1} \\
\color{mygreen}{\gamma_4} & \rule{0ex}{2.5ex}\color{mygreen}{A_1^{+-}} & \color{mygreen}{1} & \color{mygreen}{1} &
\color{red}{\sigma_{4i}} & \rule{0ex}{2.5ex}\color{red}{T_1^{--}} & \color{red}{i} & \color{red}{2} \\
\color{mygreen}{\gamma_4\gamma_5} & \rule{0ex}{2.5ex}\color{mygreen}{A_1^{-+}} & \color{mygreen}{1} & \color{mygreen}{2} &
\color{red}{\epsilon_{ijk}\sigma_{jk}} & \rule{0ex}{2.5ex}\color{red}{T_1^{+-}} & \color{red}{i} & \color{red}{1} \\ \hline 
\end{array}
$
\end{center}
\caption{Octahedral symmetries of the spinor bilinears.  Red and green entries are twist angle invariant for charged and neutral mesons respectively as shown in Section~\protect\ref{sec:oper-with-defin}.}
\label{tab:spinorreduction}
\end{table} 

\section{Link Contribution to Group Structure}
\label{sec:link-contr-group}
\begin{figure}[h]
\center{\includegraphics[width=.8\textwidth]{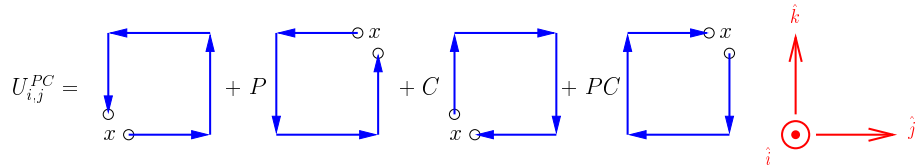}}
\caption{$PC$-symmetrized basis elements for gauge field links.}
\label{fig:upc}
\end{figure}

Define the gauge link part of our operators in equation (\ref{eq:1}) by
\begin{equation}
U_{j,k}(x)\equiv
U_j(x)U_k(x+\unit{j})U_{-j}(x+\unit{j}+\unit{k})U_{-k}(x+\unit{k})\
,
\end{equation}
where $j,k = \pm 1,\pm 2,\pm 3$; $j \ne k$.  The parity and charge
conjugation contributions due to this link structure can then be taken
into account by defining the $PC$-adapted superpositions,
\begin{equation}
U^{PC}_{i,j} = U_{j,k} + P U_{-j,-k} + C U_{k,j} + PC U_{-k,-j}\ ,
\end{equation}
where $k$ is defined via $\hat{k}=\hat{i}\times\hat{j}$ and we have
suppressed an overall normalization factor of $1/2$.
Diagrammatically, $U^{PC}_{i,j}$ is shown in Figure~\ref{fig:upc}.
For fixed $PC$ the space spanned by $U^{PC}_{i,j}$ is invariant
under~$\Ogroup$ and will generate a representation $U^{PC}$ of the
group.  A basis for the six-dimensional space may be found by
restricting $(i,j)$ to $\{(1,2),(1,3),(2,3),(2,1),(3,1),(3,2)\}$ and
this will in turn give the particular matrix representation
$\rep{U^{PC}}{(i,j)}{(m,n)}{\elem}$.  The character table for $U^{PC}$
giving $\character^{(U^{PC})}(\class)$ is shown in Table~\ref{tab:upcchartable}.
\begin{table}[h]
\begin{center}
$
\begin{array}{|ccccc|} \hline
E & 3C_4^2 & 8C_3 & 6 C_4 & 6 C_2 \\ \hline
6 & 2P & 0 & 0 & (P+1)C \\ \hline
\end{array}
$
\end{center}
\caption{Character table for representation $U^{PC}$ induced by $U^{PC}_{i,j}$.}
\label{tab:upcchartable}
\end{table}
Use of Formula (\ref{eq:2}) reduces $U^{PC}$ to irreps $\irrep^{PC}$
of $\Ogroup^{PC}$ as shown in Table~\ref{tab:upcreduction}.
\begin{table}
\begin{center}
$
\begin{array}{|cc|ccccc|} \hline
P & C & \rule{0ex}{2.5ex}A_1^{PC} & A_2^{PC} & E^{PC} & T_1^{PC} & T_2^{PC} \\ \hline
+ & + & 1 & 0 & 1 & 0 & 1 \\
+ & - & 0 & 1 & 1 & 1 & 0 \\
- & + & 0 & 0 & 0 & 1 & 1 \\
- & - & 0 & 0 & 0 & 1 & 1 \\ \hline
\end{array}
$
\end{center}
\caption{Reduction of $U^{PC}$ to irreps $\irrep^{PC}$ of
  $\Ogroup^{PC}$.}
\label{tab:upcreduction}
\end{table}

The symmetrized link fields corresponding to this reduction are
 
\begin{equation}
\link^{\irrep^{PC}}_{\irrepi}(x)= \sum_{(i,j)}\frac{\sum_{\elem \in \Ogroup}
  \rep{U^{PC}}{(i,j)}{(m,n)}{\elem}
  \rep{\irrep}{\irrepi}{\irrepj}{\elem}^*}
{\left[\frac{\order{\Ogroup}}{\irrepdim_\irrep}
    \sum_{\elem \in \Ogroup}
    \rep{U^{PC}}{(m,n)}{(m,n)}{\elem}
    \rep{\irrep}{\irrepj}{\irrepj}{\elem}^*\right]^{\frac{1}{2}}}U^{PC}_{i,j}(x)
\ ,
\end{equation}
where $(m,n)$ and $\irrepj$ are chosen so the denominator does not
vanish and the accessible $\irrep^{PC}$ are chosen from
Table~\ref{tab:upcreduction}.  (See Ref. \cite[page
74]{Cornwell:1997ke}.) Here $\rep{\irrep}{\irrepi}{\irrepj}{\elem}$ is
our matrix representation for the irrep $\irrep$. The
$\link^{\irrep^{PC}}_{\irrepi}(x)$ transform as the row $\irrepi$ of
irrep $\irrep^{PC}$ and are uniquely identified by them.

\section{Total Representation Reduction and Operator Construction}
\label{sec:total-repr-reduct}
\subsection{Reduction of Direct Product Representations}
Having classified the spinor and link components of our operators into
irreps of $\Ogroup^{PC}$, it remains to combine them and reduce all
possible direct product representations.  Parity and charge
conjugation for the product representations are given by
\begin{equation}
\begin{array}{ccc}
P=P_\diraclower P_\linklower & \hspace{1em} & C=C_\diraclower C_\linklower\ ,
\end{array}
\label{eq:3}
\end{equation}
where $\diraclower$ corresponds to the fermionic part and $\linklower$
corresponds to the gauge part.  Noting that the character of
$\irrepone\otimes\irreptwo$ satisfies
$\character^{(\irrepone\otimes\irreptwo)}(\class)=\character^{(\irrepone)}(\class)\character^{(\irreptwo)}(\class)$
and using Formula~~(\ref{eq:2}), one may reduce each direct product
into irreps of $\Ogroup$. As such the $384$-dimensional representation
generated by the space spanned by $M_{j,k,a,b}(t)$ may be reduced into
irreps of $\Ogroup^{PC}$ as shown in Table~\ref{tab:totalreduction}.
\begin{table}[h]
\begin{center}
$
\begin{array}{cccc}
\begin{array}{|c|ccccc|}
\hline
\otimes & A_1 & A_2 & E & T_1 & T_2 \\ \hline
A_1 & A_1 & A_2 & E & T_1 & T_2 \\
T_1 & T_1 & T_2 & T_1 \oplus T_2 & A_1 \oplus E \oplus T_1 \oplus T_2
& A_2 \oplus E \oplus T_1 \oplus T_2 \\
\hline
\end{array} &
\Longrightarrow
&
\begin{array}{|c|rrrr|}
\hline
\irrep\backslash PC & ++ & +- & -+ & -- \\ \hline
A_1 & 4 & 4 & 6 & 2 \\
A_2 & 4 & 4 & 2 & 6 \\
E & 8 & 8 & 8 & 8 \\
T_1 & 12 & 12 & 10 & 14 \\
T_2 & 12 & 12 & 14 & 10 \\ \hline
\end{array}
\end{array}
$
\end{center}
\caption{Reduction of representation generated by span of $M_{j,k,a,b}(t)$ into $\Ogroup^{PC}$ irreps.}
\label{tab:totalreduction}
\end{table}
Note that all possible lattice irreps are accessible, not just the
$A_1^{PC}$ and $T_1^{PC}$ of simpler local operators.

\subsection{Operator Construction Using Clebsch-Gordan Coefficients}
The construction of the lattice-symmetrized operators themselves is
accomplished by combining the spinor and link operator components
$\dirac^{\irrep^{PC},\irrepcopy}_\irrepi$ and
$\link^{\irrep^{PC}}_\irrepi$ using the same formulae for parity and
charge conjugation given in Equation (\ref{eq:3}) and by combining the
irreps of $\Ogroup$ using Clebsch-Gordan (C-G) coefficients for
$\irrepone\otimes\irreptwo$.  The simplicity of the possible direct
product representations allows one to use
\begin{equation}
\left(\begin{array}{cc|c}
\irrepone & \irreptwo & \irrep \\
\irreponei & \irreptwoi & \irrepi
\end{array}
\right)
= \frac{\sum_{\elem \in \Ogroup}
  \rep{\irrepone}{\irreponei}{\irreponej}{\elem}
  \rep{\irreptwo}{\irreptwoi}{\irreptwoj}{\elem}
  \rep{\irrep}{\irrepi}{\irrepj}{\elem}^*}
{\left[\frac{\order{\Ogroup}}{\irrepdim_\irrep}
    \sum_{\elem \in \Ogroup}
    \rep{\irrepone}{\irreponej}{\irreponej}{\elem}
    \rep{\irreptwo}{\irreptwoj}{\irreptwoj}{\elem}
    \rep{\irrep}{\irrepj}{\irrepj}{\elem}^*\right]^{\frac{1}{2}}}
\end{equation}
to determine the C-G coefficients.  (See Ref. \cite[page
74]{Cornwell:1997ke}.)  Here $\irrepj$, $\irreponej$, and $\irreptwoj$
are chosen so the denominator does not vanish.  The
lattice-symmetrized operators, $M^{\irrep^{PC}}_\irrepi(t)$, are then
given by
\begin{equation}
\mesonop_\irrepi^{\irrep^{PC},\irrepdirac^{P_\diraclower C_\diraclower}, \irrepcopy_\diraclower, 
\irreplink^{P_\linklower C_\linklower}}(t) 
={\displaystyle\sum_\vecx\sum_{\irrepdiraci, \irreplinki}}
\left(\begin{array}{cc|c}
\irrepdirac & \irreplink & \irrep \\
\irrepdiraci & \irreplinki & \irrepi
\end{array}
\right)
\spinorbar(x)\dirac^{\irrepdirac^{P_\diraclower C_\diraclower}, \irrepcopy_\diraclower}_{\irrepdiraci}
\link^{\irreplink^{P_\linklower C_\linklower}}_{\irreplinki}(x)\spinor(x)
\ ,
\end{equation}
where the allowed irreps for the spinor bilinear and link components
are determined by Tables~\ref{tab:spinorreduction}
and~\ref{tab:upcreduction}.  These fix $P$ and $C$ for the operator
while the irreps $\irrep$ of $\Ogroup$ are those allowed by the C-G
series in Table~\ref{tab:totalreduction}.  Dirac indices on $\dirac$
and color indices on $\link$ and both indices on the spinors have been
suppressed.

\section{Twisted Mass Invariant Operators}
\label{sec:twist-mass-invar}
\subsection{Operators with Definite Parity}
\label{sec:oper-with-defin}
Consider tmLQCD for a quark doublet.  A change of basis (by twist
angle $\omega$) gives
\begin{equation}
\left(\begin{array}{c} u^\prime \\ d^\prime \end{array}\right)
= \exp\left(i\omega\gamma_5\tau_3\right)
\left(\begin{array}{c} u \\ d \end{array}\right)
\ ,
\end{equation}
which leads to mixing among some, but not all, quark bilinears.  For
example,
\begin{equation}
\bar{u}^\prime\gamma_i d^\prime
= \bar{u}\gamma_i d\cos\omega + \bar{u}\gamma_i\gamma_5 d\sin\omega
\end{equation}
couples to both vector and axial mesons unless $\omega$ is carefully tuned, but
\begin{equation}
\bar{u}^\prime\sigma_{4i} d^\prime = \bar{u}\sigma_{4i} d
\end{equation}
couples to the vector and never to the axial.  Explicit calculation
shows that tmLQCD splits the results of
Section~\ref{sec:spin-contr-group} into two sets according to flavor.
Structures in {\color{red} red} in Table~\ref{tab:spinorreduction} are
only independent of twist angle for charged mesons, while
{\color{mygreen} green} structures are only independent of twist angle
for neutral mesons.  Despite these restrictions, one finds that {\em
  all $\Lambda^{PC}$ combinations can still be obtained by factoring
  in some nontrivial link structure}.  Combining the twist-invariant
spinor structures from Table~\ref{tab:spinorreduction} with the link
structure in Table~\ref{tab:upcreduction} results in operators of the
irreps outlined in Table~\ref{tab:twistreduction} being accessible.
\begin{table}[h]
\begin{center}
{\color{red}
\begin{tabular}{|c|cccc|}
\multicolumn{5}{c}{\color{black}{charged mesons}*} \\
\hline
$\Lambda\backslash PC$ & ++ & +- & -+ & -- \\
\hline
$A_1$ & 3 & 1 & 3 & 1 \\
$A_2$ & 1 & 3 & 1 & 3 \\
$E$   & 4 & 4 & 4 & 4 \\
$T_1$ & 5 & 7 & 5 & 7 \\
$T_2$ & 7 & 5 & 7 & 5 \\
\hline
\end{tabular}
}
\hspace{1cm}
{\color{mygreen}
\begin{tabular}{|c|cccc|}
\multicolumn{5}{c}{\color{black}{neutral mesons}} \\
\hline
$\Lambda\backslash PC$ & ++ & +- & -+ & -- \\
\hline
$A_1$ & 1 & 3 & 3 & 1 \\
$A_2$ & 3 & 1 & 1 & 3 \\
$E$   & 4 & 4 & 4 & 4 \\
$T_1$ & 7 & 5 & 5 & 7 \\
$T_2$ & 5 & 7 & 7 & 5 \\
\hline
\end{tabular}
}
\end{center}
\caption{Twist invariant irreps for charged and neutral mesons respectively.  *It is to be noted here that charged mesons are really only eigenstates
  of g-parity with eigenvalue $G=C(-1)^I$, but we use $C$ for simple
  correspondence with the previous arguments.}
\label{tab:twistreduction}
\end{table}

\subsection{Preserving Parity in the Correlators}
Building states of definite $\Lambda^{PC}$ might seem insufficient for
tmLQCD, since the theory does not preserve $P$.  However, tmLQCD does
conserve a ``twisted parity'', $\tilde P$, defined by the product of
$P$ and $(\omega\to-\omega)$.  Therefore, states of definite twisted
parity are automatically preserved in any correlation function.  Since
our quark bilinears are {\em independent} of $\omega$, $P$ and $\tilde
P$ are equivalent for these bilinears.

\section*{Acknowledgements}
This work was supported in part by the Natural Sciences and Engineering
Research Council of Canada, the Canada Foundation for Innovation, the
Canada Research Chairs Program and the Government of Saskatchewan.


\end{document}